\documentclass[aps,pra,reprint,superscriptaddress]{revtex4-1}

\usepackage[]{hyperref}
\hypersetup{colorlinks=true}
\usepackage{graphicx}
\usepackage{dcolumn}
\usepackage{bm}
\usepackage{helvet}
\usepackage{amssymb}
\usepackage{amsfonts}
\usepackage{color}
\usepackage{epstopdf}
\usepackage{ulem}

\newcommand{\mean}[1]{\ensuremath{\left\langle #1 \right\rangle}}

\usepackage{xcolor}

\newcommand{\pref}[1]{(\ref{#1})}
\newcommand{\eref}[1]{Eq.~\pref{#1}}
\newcommand{\fref}[1]{Fig.~\ref{#1}}

\usepackage{etex}
\reserveinserts{18}
\usepackage{morefloats}

\begin{document}

\author{Juan Polo}
\affiliation{Quantum Systems Unit, Okinawa Institute of Science and Technology Graduate University, Onna, Okinawa 904-0495, Japan}
\affiliation{Univ. Grenoble Alpes,  CNRS, LPMMC, F-38000 Grenoble, France}

\author{Piero Naldesi}
\affiliation{Univ. Grenoble Alpes, CNRS, LPMMC, F-38000 Grenoble, France}

\author{Anna Minguzzi} 
\affiliation{Univ. Grenoble Alpes, CNRS, LPMMC, F-38000 Grenoble, France}

\author{Luigi Amico}
\affiliation{Dipartimento di Fisica e Astronomia, Via S. Sofia 64, 95127 Catania, Italy}
\affiliation{Centre for Quantum Technologies, National University of Singapore, 3 Science Drive 2, Singapore 117543, Singapore}
\affiliation{MajuLab, CNRS-UNS-NUS-NTU International Joint Research Unit, UMI 3654, Singapore}
\affiliation{CNR-IMM \& INFN-Sezione di Catania, Via S. Sofia 64, 95127 Catania, Italy}
\affiliation{LANEF \textit{'Chaire d'excellence'}, Univ. Grenoble-Alpes \& CNRS, F-38000 Grenoble, France}

\title{Exact results for persistent currents of two bosons in a ring lattice \\
}

\begin{abstract}
  We study the ground state of two interacting bosonic particles confined in a ring-shaped lattice potential
  and subjected to a synthetic magnetic flux. The system is described by the Bose-Hubbard model  and solved  exactly through a plane-wave Ansatz of the  wave function. We
  obtain energies and correlation functions of the system both for repulsive and attractive interactions. In contrast with the one-dimensional continuous theory described by the Lieb-Liniger model, in the lattice case we prove that the center of mass of the two particles is coupled with its relative coordinate. Distinctive features clearly emerge in the persistent current of the system. While for repulsive bosons the persistent current displays a periodicity given by
  the standard flux quantum for any interaction strength, in the attractive case the flux quantum becomes fractionalized in a manner that depends on the interaction. We also study the density after the long time expansion of the system which provides
  an experimentally accessible 
  route to detect persistent currents in cold atom settings. Our results can be used to benchmark approximate schemes for the many-body problem.
\end{abstract}

 \maketitle

\section{Introduction}

Bosonic particles confined in a ring geometry can be realized with different quantum technological platforms, ranging from superconducting circuits \cite{krantz2019quantum} to circuit quantum electrodynamics (QED)  \cite{blais2007quantum}, and to cold atoms  \cite{Morizot2006, franke2007optical, wright2013driving, gauthier2016direct, muldoon2012control, Boshier_painting}. 
There are many reasons explaining the relevance of such systems for the physics community. Firstly, the periodic boundary conditions provide a textbook route to emulate quantum systems with strict translational symmetry and to simplify the access to the large number of particles ($N$) regime of the system (thermodynamic limit)  \cite{fisher1972scaling}. On the other hand, systems with closed spatial architectures provide the simplest instance of quantum circuits that are able to sustain non-trivial current states that can be used in quantum technology to construct quantum devices and quantum sensors with enhanced performances  \cite{acin2018roadmap, ragole2016interacting}. Here, we focus on the persistent currents, which are bosonic current states originating from the coherence of the system \cite{ambegaokar1990coherence, 1983PhLA...96..365B, 1990PhRvL..64.2074L, 2009Sci...326..244B}. 
Such specific current states of thermodynamic nature can be imparted to the bosons by a real or synthetic 
magnetic field \cite{dalibard2011colloquium}.
Persistent currents have been of defining importance in mesoscopic physics \cite{imry2002introduction}. Recently, persistent currents have been attracting upsurged interest, especially in the cold-atom 
community as they grant an enhanced flexibility and control over the physical conditions of the system \cite{Ramanathan2011,Ryu2013,eckel2014hysteresis, amico2014superfluid,aghamalyan2015coherent,aghamalyan2016atomtronic,aghamalyan2013effective,Mathey_Mathey2016,haug2018readout,PhysRevA.84.053604,PhysRevLett.113.025301,Turpin:15,Amico_Atomtronics}.

Bose statistics enables  a multi-particle interaction that can make interacting bosonic systems difficult to handle. In this respect, it is  instructive to consider the interplay between the Bose gas field theory (point like interaction) and the Bose-Hubbard model, providing two celebrated schemes describing many-body bosonic theories. The homogeneous Bose gas in one spatial dimension, described by the Lieb-Liniger model, is a continuous field theory that is integrable by Bethe Ansatz: the many-body scattering can be factorized in two-particle scatterings where the scattering is 'non-diffractive' \cite{korepin1997quantum}. In contrast, the Bose-Hubbard model, introduced by  Haldane as a lattice regularization of the Bose-gas field theory, is not integrable and the scattering is diffractive \cite{haldane1980solidification, choy1982failure, choy1980some}. The two theories are equivalent in a dilute limit of very few particles per lattice site. In such a limit, in which  multiple occupancy
is loosely speaking
demoted, the Bose-Hubbard dynamics becomes integrable. Such considerations can be expressed quantitatively through coordinate Bethe Ansatz. Indeed, it was demonstrated that the Schr\"odinger equation arising from Bose-Hubbard model cannot be solved by coordinate Bethe Ansatz if the probability of having more than two particles in the same site is non vanishing \cite{choy1982failure}. Thus,  the two-particle Bose-Hubbard model is analytically accessible.  Despite the simplicity of the system, the $N=2$ Bose-Hubbard Model (2BHM) has been demonstrated to provide a very useful case study to decode some of the general features of the many-body theory. Recently, the 2BHM has been employed to answer relevant questions concerning cold atoms confined in optical lattices \cite{valiente2008two, valiente2009scattering, valiente2008quantum, valiente2010lattice} 
Dynamical effects of bosonic pairs in a one dimensional lattice, both in the attractive and repulsive case, have been also analysed \cite{Piero_2014}.

In this work, we will employ the 2BHM to the  persistent currents in a ring-shaped lattice potential by exact means. 
To this end, we use 
the exact expression of the two-body wave function in the presence of an external gauge field. We calculate relevant observables and correlation functions characterizing the system. We discuss 
the differences between our results and those coming from the continuous Bose gas field theory. 

The article is outlined as follows: in Sec.~II, we present the model system together with the expressions for the exact spectrum and correlation functions, in Secs.~\ref{ResultsPositive} and \ref{ResultsNegative}, we discuss the results for positive and negative interactions respectively, and finally in Sect.~\ref{conclusions}, we draw our conclusions.

\section{Model system and observables}
\label{ModelMethods}
We model the system of interacting bosons trapped in a $L$-site  lattice ring under the presence of synthetic gauge field using the Bose Hubbard model: 
\begin{eqnarray}
\label{eq:HBH}
{H}(\Omega, U)&=& {K}(\Omega) + {V}(U) \\
{K}(\Omega) 
&=& -J_0 \sum_{j=1}^L
\left(e^{i 2 \pi \Omega / L} \:{b}_j^\dagger {b}_{j+1} + \text{h.c.} \right)\, , \nonumber \\
{V} (U)&=& +\frac{U}{2}\sum_{j=1}^L{n}_j\left({n}_j-1\right) \nonumber
\end{eqnarray}
where ${b}_j$ is the bosonic annihilation operator and ${n}_j={b}^\dagger_j {b}_j$ is the local number operator for the site $j$. The parameters $U$ and $J_0$ account for the strength of the on-site-interactions and tunneling amplitude respectively. We assume that the size of the system is sufficiently large such that the Peierls substitution becomes well defined  \cite{Peierls_1933}. 

In the next sections, we will discuss the properties of the system for both positive and negative interactions $U$. 
Since $V(-U)=-V(U)$ and $\mathcal{K}(\Omega)= - \mathcal{K}\left(\Omega+ L/2 \right)$, the following simmetries connect the two cases 
\begin{eqnarray}
H(\Omega, U)&=&{K}\left(\Omega+ L/2 \right) - {V}(-U)] \\
&=& - {H}(\Omega+ L/2, -U)\; ;
\end{eqnarray}
in addition:
\begin{equation}
H(\Omega, U)=-H(\Omega, -U)
\end{equation}
 The current operator, which is the most relevant observable to describe persistent currents, reads:
\begin{eqnarray}
I(\Omega) = i J_0 \sum_j \left( e^{2\pi i \Omega/L} b^\dagger_j b_{j+1} - \text{h.c.} \right) \;.
\label{current}
\end{eqnarray}
The persistent current can also be obtained through thermodynamic potentials and, in particular, at zero temperature it is given by: 
\begin{equation}
\langle I(\Omega) \rangle = - \frac{\partial E}{\partial \Omega}
\label{persistent}
\end{equation}
where $E$ is the ground state energy of the system. 

A classic result in the field was obtained by Leggett \cite{nanoelectronics1991dk}. By resorting to the analogy of particles moving in a magnetic field and using the Bloch theorem for  particles in a periodic potential, it can be demonstrated that the energy of the many-body system displays a periodicity in $\Omega$ that is
fixed by the  elementary flux quantum of the system. Therefore, due to Eq.~(\ref{persistent}), the persistent current is also  a periodic function of $\Omega$
with the same periodicity. This result holds for any local two-body interaction.  In the next sections, the Leggett results will be analysed for the 2BHM.   

In the limit of small filling fractions $\nu=N/L=D \Delta$, with $D=N/(L \Delta)$, $\Delta$ being the density and the lattice spacing respectively, the  one dimensional Bose-Hubbard model can be mapped into  the integrable Bose gas field theory or, in first quantization,  the Lieb-Liniger model\cite{amico2004universality, dutta2015non}. Here, we  keep the dimensionless coupling strength of the Lieb-Liniger model,  $\gamma=mg/\hbar^2 D$ constant, being   $g$ the delta interaction coupling constant of the continuous model and $m$ the particles mass. The  parameters of the lattice and continuous theories are related by  $U=g/\Delta$ and $J_0=\hbar^2/2m \Delta^2$, yielding $\gamma= \nu^{-1}(U/J_0)$. Hence, by increasing the number of lattices sites $L$ at fixed $N$,  the interaction energy $U/J_0$ should be decreased in order  the Bose gas limit is achieved. 
 We note that, since the continuous rotational symmetry of the Lieb-Liniger model is reduced to a discrete one,  
in the lattice  the current  operator is generically distinct from the angular momentum. In particular,  $I(\Omega)$  in Eq. (\ref{current}) does not commute with the  Hamiltonian in Eq.(\ref{eq:HBH}) (interaction term $V(U)$). In a dynamical protocol, therefore, the current would not be  conserved \cite{zotos2004transport}. Nevertheless, $\langle I(\Omega)\rangle$  will be denoted  as 'persistent current' in analogy of the current states of  the continuous theory (see also \cite{Arwas_2017}).
Important insights on the current states of the system can be obtained by the current's fluctuations
\begin{eqnarray}
\Delta I (\Omega) = \sqrt{ \mean{I(\Omega)^2} - \mean{I(\Omega)}^2 }
\end{eqnarray}
In cold atom settings, the current state manifests itself in the time-of-flight expansion (TOF) images, {\it i.e.} the particle density  after  the condensate is released from the confining potential. The long time density pattern can be calculated through the momentum distribution at the instant at which the trap is opened, according to 
\begin{eqnarray}
n(\mathbf{k}) &=& |w(\mathbf k)|^2 \sum_{j, l} e^{i \mathbf{k}\cdot(\mathbf{x}_j-\mathbf{x}_l)} \langle b^\dagger_j b_l\rangle \;,
\end{eqnarray}
where $\mathbf{x}_j$ is the position of the lattice sites in the plane of the ring and $w(\mathbf k)$ are the Fourier transforms of the Wannier functions.

In the following, we will obtain exact expressions both for the spectrum and correlation functions of the 2BHM, granting us access to the relevant observables describing the persistent currents of the system.

\subsection{The Bose-Hubbard model in the two-particle sector}

The 2BHM model is exactly solvable {\em \`a la} coordinate Bethe Ansatz  \cite{Piero_2014}. As we shall see, this approach allows us to deal with the persistent current of the system exactly, in the case in which the ring is exposed to an effective magnetic field. 
A general two-boson state can be written as:
 \begin{equation}
 \left|\phi\right>=\sum_{j, k=1}^L\phi_{jk}\hat{b}_j^\dagger \hat{b}_k^\dagger\left|0\right>	\, , 
 \end{equation}
whith $\phi_{jk}$ being symmetric under the exchange of $j$ and $k$ and $\langle \phi \left|\phi\right>=1$.
 The Schr\"odinger equation $H\left|\phi\right>=E\left|\phi\right>$ reads:
\begin{eqnarray}
\label{eq:Schrodinger}
F_{jk}&=&\left(E-U\delta_{jk}\right)\phi_{jk}
\end{eqnarray}
where is given by $F_{jk}\doteq -J\left(\phi_{j+1, k}+\phi_{j, k+1}\right)-J^*\left(\phi_{j-1, k}+\phi_{j, k-1}\right)$.

The solution of Eq.(\ref{eq:Schrodinger}) is obtained with a plane-wave Ansatz for $\phi_{jk}$, which is certainly correct in the non interacting limit  $U=0$:
 \begin{eqnarray}\label{BA}
 \phi_{jk}&=\left[a_{12}e^{i\left(p_1j+p_2k\right)}+a_{21}e^{i\left(p_1k+p_2j\right)}\right]\vartheta(j-k)+ \nonumber\\
 &+\left[a_{12}e^{i\left(p_1k+p_2j\right)}+a_{21}e^{i\left(p_1j+p_2k\right)}\right]\vartheta(k-j)	\, .
 \end{eqnarray}

Equation (\ref{eq:Schrodinger}) for $j\neq k$ is solved using: 
\begin{eqnarray}
E&=& \langle\phi|\hat{H}|\phi\rangle = -2J\left(\cos p_1+\cos p_2\right) \nonumber\\
&=&-4J\cos\left( \frac{P}{2} + \frac{2\pi\Omega}{L} \right)\cos\left( p \right)	\, .
\end{eqnarray}
 By introducing the center-of-mass and relative coordinates for both space and momentum coordinates:
\vspace{-0.25cm}
 \begin{eqnarray}
 X=\frac{j+k}{2} \ \ \ &&\ \ \ 
 x=j-k\\
 P=p_1+p_2 \ \ \ &&\ \ \ 
 p=\frac{p_1-p_2}{2}	\, .
 \end{eqnarray}
\vspace{-0.25cm}
we can rewrite equation~(\ref{BA}) as:
\vspace{-0.25cm}
\begin{equation}\label{eq:BAXx}
\phi_{Xx}=a_{1, 2} e^{iPX}\left(e^{ip\left|x\right|}+\frac{a_{21}}{a_{12}} e^{-ip\left|x\right|}\right) \; .
\end{equation}
\vspace{-0.25cm}

The plane-wave Ansatz (\ref{BA}) is exact 
also for  interacting particles since for $U\neq 0$ the effect of the interaction can be recasted into the scattering between the two sectors, $j>k$ and $j<k$, in which the particles are non interacting. Indeed, by introducing (\ref{BA}) into \eref{eq:BAXx} for $j=k$, the ratio between the coefficients, $y(P, p)\doteq\frac{a_{21}}{a_{12}}$, can be obtained as a phase shift of the wave function:
\begin{widetext}
\begin{eqnarray}
y(P, p)
&=&
-
\frac{
\frac{U}{4 J_0}
-i\cos\left( \frac{P}{2} - \frac{2 \pi \Omega }{L} \right) \sin(p)
-2\cos(p)
\sin(\frac{P}{2}-\frac{\pi \Omega }{L})
\sin(\frac{\pi \Omega }{L})
}
{
\frac{U}{4 J_0}
+i\cos\left( \frac{P}{2} - \frac{2 \pi \Omega }{L} \right) \sin(p)
-2\cos(p)
\sin(\frac{P}{2}-\frac{\pi \Omega }{L})
\sin(\frac{\pi \Omega }{L})
} 	\, , \qquad |y\left(P, p\right)|=1
\end{eqnarray}
\end{widetext}

The allowed values of the momenta are fixed by the boundary conditions. If not otherwise stated, we assume periodic boundary conditions
\begin{equation}\label{PBC_sc}
\phi_{j, 1}= \phi_{j, L+1} \;, 
\end{equation}
leading to the following equations for the center of mass and relative momenta:
\begin{eqnarray}\label{eq:Pandp}
P_n = \frac{2\pi n}{L}\, ;\qquad (-1)^ne^{ipL}=y\left(P_n, p\right)	\, .
\end{eqnarray}
with $n=\{1, \cdots, L\}$. 

The relative momentum $p$ depends on $U/J_0$ and $\Omega$ explicitely, and can be either real or complex-valued giving rise to {\it scattering} or {\it bound} states respectively. With the chosen periodic boundary conditions, the center of mass momentum $P$ does not depend directly on $\Omega$ or on $U/J$. Alternatively, with the twisted boundary conditions $\phi_{j, 1}= e^{i\Omega} \phi_{j, L+1} $, $\Omega$ can be gauged away from the scattering matrix. In this case, $\Omega$ would have shifted the center of mass $P$ (see Appendix). 

\subsubsection{Correlation functions.}
In this section, we calculate the one-body (two-point) and two-body (four-point) correlations that are needed to map-out the observables of the system. 
The two-point correlation function is:
\begin{eqnarray}
C^{1b}_{r}&=&\mean{b_l^\dagger b_{l+r}}=\sum_m \phi_{l, m}^*\phi_{m, l+r}
\end{eqnarray} 
Resorting to the translational invariance of the system, we can set $l=0$ and $m>0$. We obtain:
\begin{eqnarray}
\label{one-body}
C^{1b}_{r}
&=&2 \mathcal{N}^2\: e^{i P \frac{r}{2}} \bigg[ \nonumber
\frac{1}{2} \csc (p) 
\Big(
\sin (p (L+r+1) -\psi )\\ \nonumber
&+& \sin (p (L-r-1) -\psi ) - \sin (p (1+r)-\psi ) \\ \nonumber
&+& \sin (p (1-r) +\psi ) + \sin (p+p r) -\sin (p-p r) \Big)\\
&+&(L-r) \cos (p r)+r \cos (p r-\psi )
 \bigg] 
\end{eqnarray} 
where $\psi\doteq\xi_{12} - \xi_{21}$ with $\xi_{i, j}$ given by $a_{12}=\mathcal{N} e^{-i \xi_{12}}$ and $a_{21}=\mathcal{N} e^{-i \xi_{21}}$ (note that, by construction, $|a_{21}/a_{12}|^2 = 1 $).

\begin{figure*}
\includegraphics[width=\linewidth]{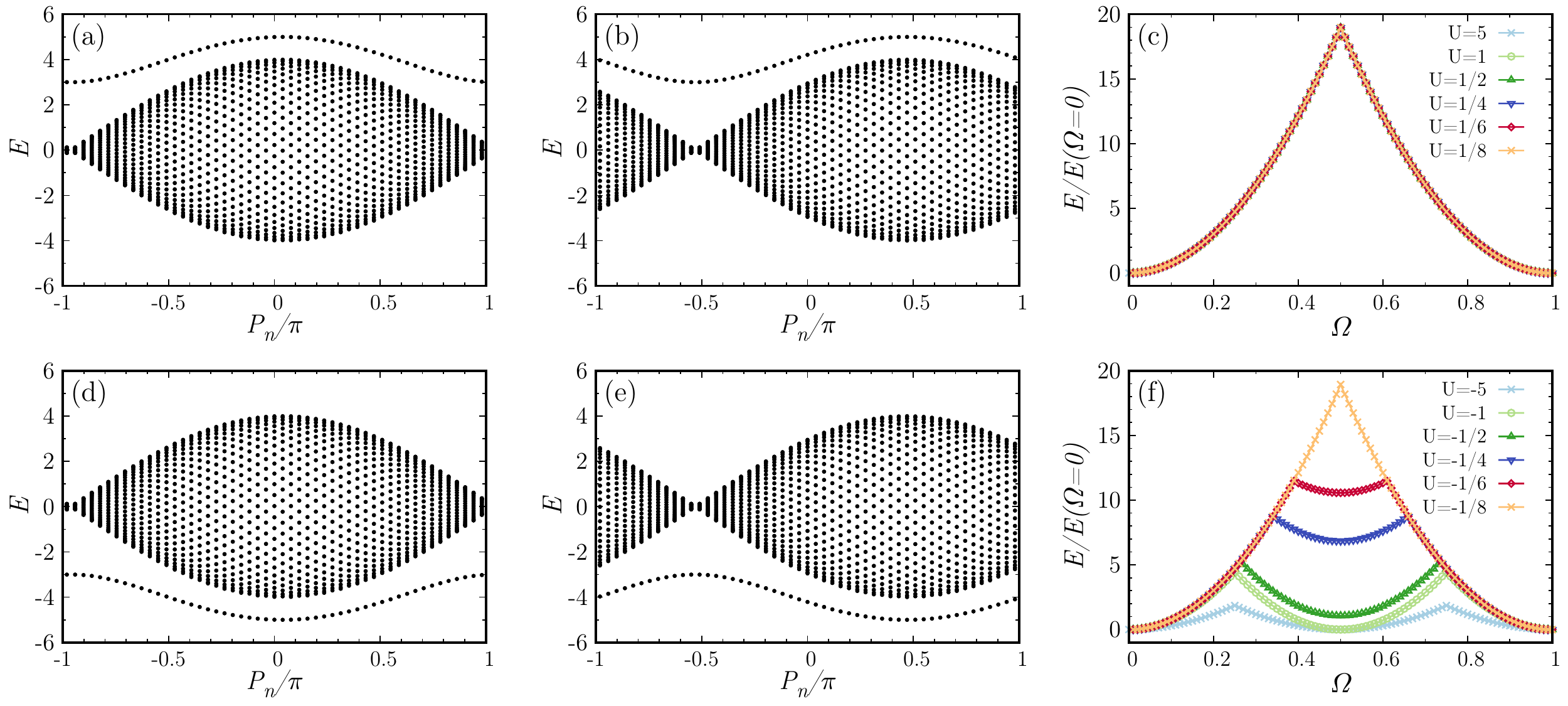}
\caption{Energy spectrum of the model for a lattice ring of $L=51$ sites. Panels (a) and (b) have $U=5$ (first row) while (d) and (e) have $U=-5$ (second row). In panels (a) and (d) the flux is set to $\Omega=0.5$ while in panels (b) and (e) is set to $\Omega=0.25L$. In panels (c) and (f) we plot the energy of the ground state as a function of the flux, rescaled by its value at $\Omega=0$, for different values of the interaction strength as indicated on the figure.}
\label{fig:spectrum_energy_vs_omega_repulsive}
\end{figure*}

The density-density correlations \mean{n_l n_{l+r}} are:
\begin{eqnarray}
\label{density-density}
C^d_{r} &=& \mean{n_l n_{l+r}} = |\phi_{l, l+r}|^2 \nonumber \\
&=& 2 \mathcal{N}^2 \left[ 1 + \text{Re}\left( e^{i\psi -2ipr} \right ) \right]
\end{eqnarray}
The connected correlation, $\mean{n_{l}n_{l+r}}-\mean{n_{0}n_{0}}$, can be calculated by subtracting $(N/L)^2$ from the density-density correlation (\ref{density-density}).

Both $C^{1b}_{r}$ and $C^d_{r}$ depend on $p$ explicitely. Therefore, the actual behavior of such quantities is substantially affected by whether the particle are attractive or repulsive. 

For attractive interactions, the lowest energy eigenstate for each center of mass wavevector has a complex-valued relative momentum, such that $p=i \alpha$, which leads to an exponential decay in the correlation functions:
\begin{eqnarray}
C^{d, \text{BOUND}}_{r}=2 \mathcal{N}^2 \left[ 1 + e^{-2 \alpha r}\cos\left( \psi \right ) \right]\end{eqnarray}
We define the scale at which the correlations decay as $\xi=1/\alpha$. This quantity defines the characteristic length of the bound state.


Finally, we calculate the pair correlation function:
\begin{eqnarray}
C^{p}_{l, j}&=&\mean{b_l^\dagger b_l^\dagger b_jb_j} = \phi_{l, l}^*\phi_{j, j} \nonumber \\
&=& 2\mathcal{N}^2 \left (
\cos\left( rP/2 \right) - i\sin\left (r P/2 \right)
\right)
\end{eqnarray}
We note that this quantity does not depend on $p$ and therefore does not distinguish between bound and scattering states.\\

\section{Results for repulsive interactions}
\label{ResultsPositive}
In this section we obtain the exact spectrum of the Hamiltonian and the persistent currents for positive $U$, thus  generalizing the results obtained  in Ref.\cite{Piero_2014}  at $\Omega=0$.

\subsection{Energy spectrum}

We start by analyzing
the spectrum and the ground-state energy of the system. We note that there are two main bands in the system (see \fref{fig:spectrum_energy_vs_omega_repulsive}(a-b)). The lowest band is characterized by real rapidities and therefore corresponds to scattering states. On top of it, we find a distinct band of bound states with complex rapidities. The two bands are separated by a finite gap, formed by the energy eigenstates with largest energy eigenvalue for each center of mass momentum $P_{n}$. Figure~\ref{fig:spectrum_energy_vs_omega_repulsive}(c) shows the ground state energy of the system as a function of the induced flux $\Omega$. Note that the interactions change the ground state independently of the magnetic field \cite{Piero_2014}. 


\begin{figure}[h!!]
\includegraphics[width=\linewidth]{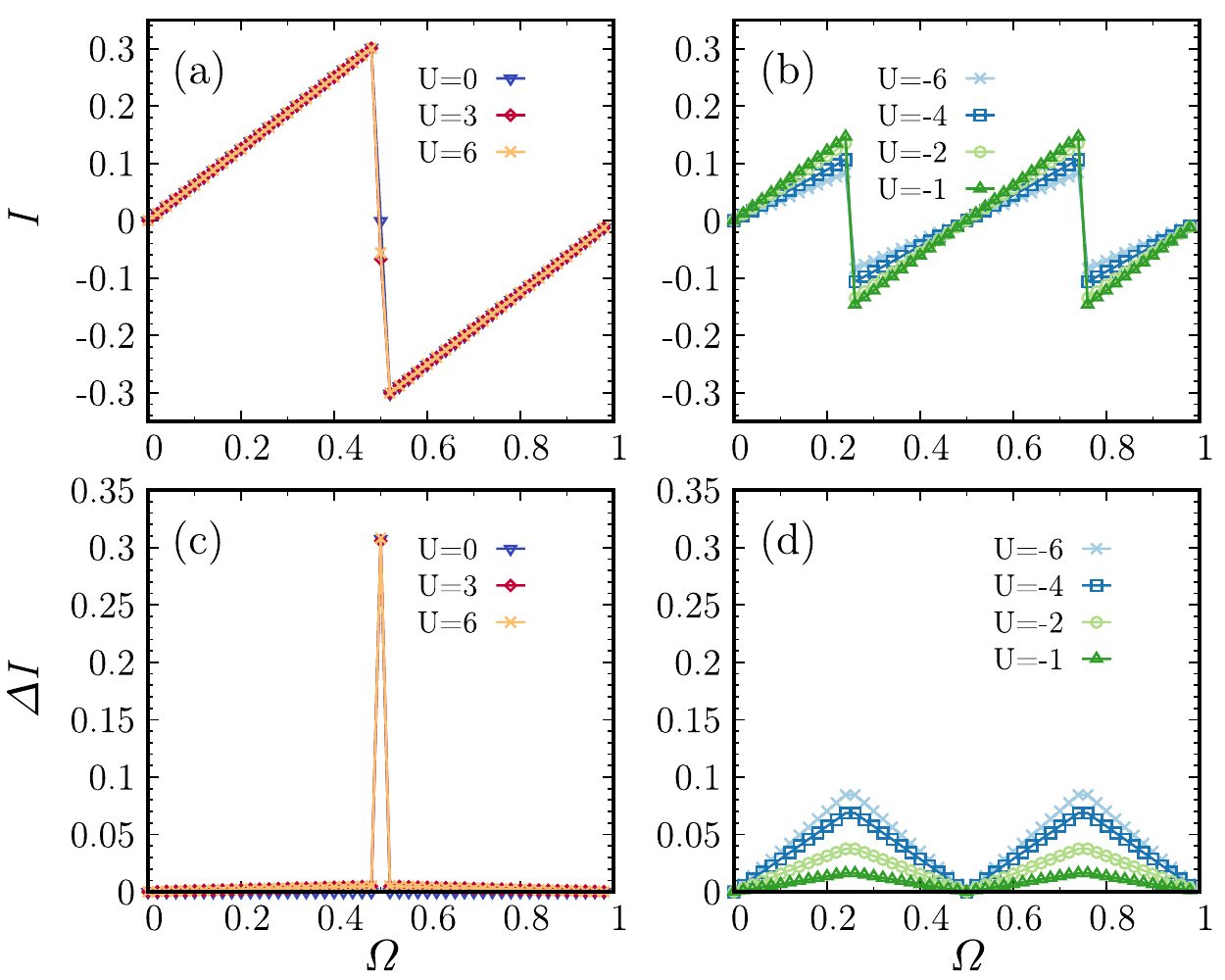}
\caption{Current (upper row) and its fluctuations (lower row) as a function of $\Omega$ for different values of the interaction strength. Left panels: repulsive interactions, right panels: attractive interactions.}
\label{fig:current_vs_omega_fluctuations}
\end{figure}

\subsection{Persistent currents}
The ground state persistent current displays the characteristic saw-tooth dependence on the synthetic
magnetic field. The jumps of the persistent currents (from clockwise to anticlockwise) and the slope of the saw-tooth are determined by the flux quantum of the system  \cite{nanoelectronics1991dk}. In the repulsive case, the current jumps occur at values that are independent of the particle number and interaction strength (see \fref{fig:current_vs_omega_fluctuations}(a)).
Moreover, the slopes of the saw-tooth behaviour of the current  are also independent on the interaction $U$. This scenario indicates that the flux quantum is a fixed quantity (independent of $N$ and $U$). 

At $\Omega=1/2$ where the jump in the current occurs, the fluctuations $\Delta I$ have the largest value. This behaviour indicates that the system is in a superposition among two states, corresponding to the $0-th$ and $1-st$ energy minima.

\begin{figure*}
\includegraphics[width=\linewidth]{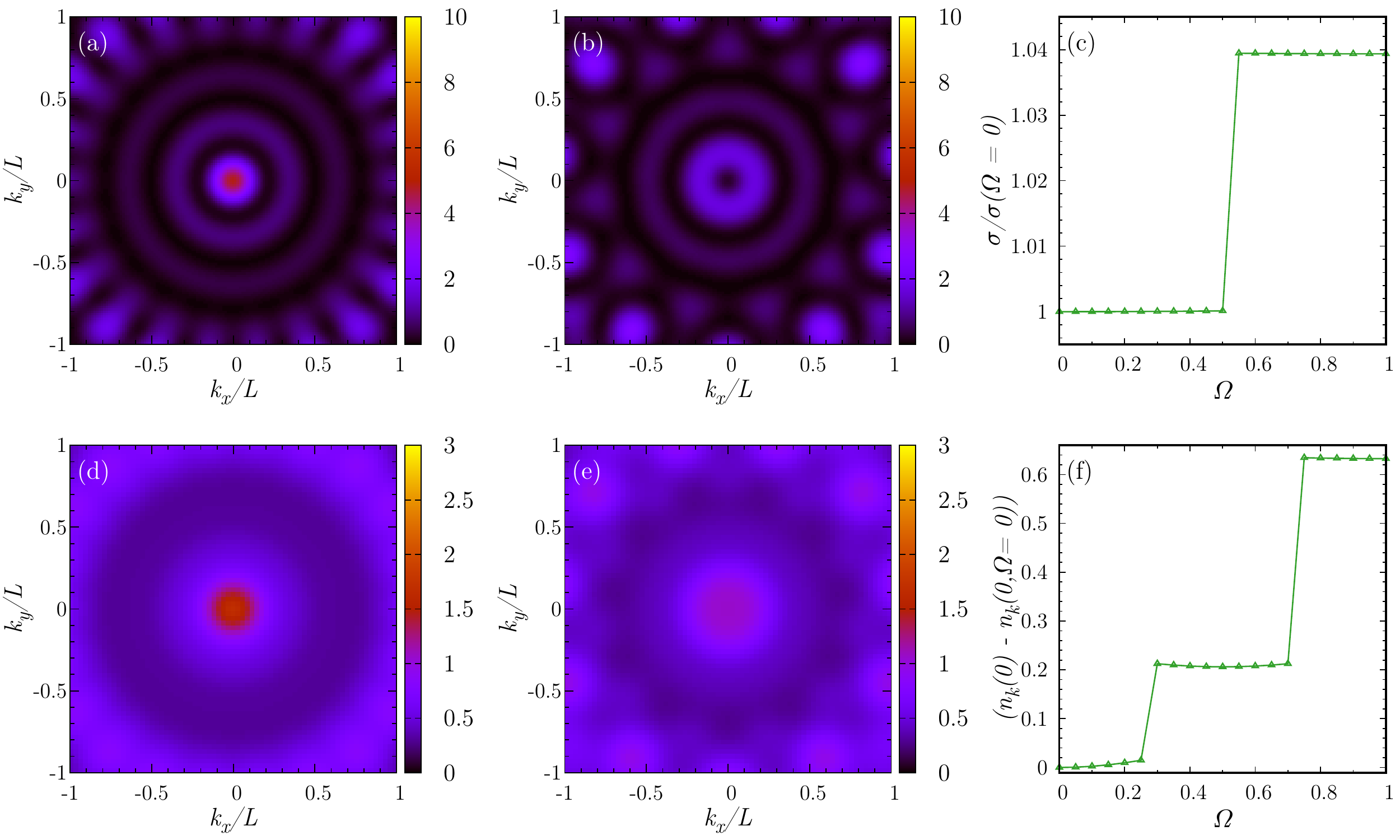}
\caption{Time of Flight density plots are displayed in panels (a), (b), (d) and (e) for a chain of $L=11$ sites. Panel (c) and (f) show the TOF central value, $n_k(\mathbf{k} = 0)$, and the TOF central peak width, $\sigma$, respectively. Top panels show the repulsively interacting case with $U=5$ and lower panels the attractively interacting one with $U=-5$. Density plots correspond to a) $\Omega=0.0$, b) $\Omega=0.8$, d) $\Omega=0.0$ and e) $\Omega=0.8$. To highlight the TOF features we set $|w(\mathbf k)|^2 = 1$ in (a-b) and (d-e). The dispersion of the interference pattern (c), with $\sigma^2 = \int d\mathbf{k}\, \mathbf{k}^2 n(\mathbf{k})$, is calculated with the re-scaled Wannier function $|w(\mathbf k)|^2 = \exp\left((k_x^2 + k_y^2)/(L/2)^2\right)$ and it is normalized by $\sigma(\Omega=0)$.}
\label{fig:TOF_L_11}
\end{figure*}

The current states in the system arise in the long-time expansion of the density of the cold atom gas after releasing the confining potential. For $\Omega=0$, the interference pattern displays a marked peak at $\mathbf{k}=0$. For $\Omega$ larger than the first degeneracy point in the ground state energy, $\langle n(\mathbf{0}) \rangle$ is depressed, with a characteristic ring-shape symmetry (see \fref{fig:TOF_L_11}(a-c)). 
It has been demonstrated that the radius of such ring-shape feature in the expansion increases with $\Omega$ in quantized steps \cite{moulder2012quantized} thus, we calculate the TOF dispersion $\sigma^2 = \int d\mathbf{k}\, \mathbf{k}^2 n(\mathbf{k})$ to characterize such characteristic increase (see \fref{fig:TOF_L_11}(c)).

\section{Results for attractive interactions}
\label{ResultsNegative}
In this section we consider the case  $U<0$ and present the excitation spectrum
 and the ground-state persistent current as a function the frequency $\Omega$. 


\subsection{Energy spectrum}

We first focus on the energy spectrum of the two particle system (see \fref{fig:spectrum_energy_vs_omega_repulsive}(d-e)).
The results at $\Omega=0$, are in agreement with the $N$-particle case obtained through the study of the dynamical structure factor \cite{Naldesi2019}.

For  non-zero  $\Omega$ the spectrum is displaced, with a maximum displacement given by the maximum momentum allowed by the lattice. This is due to the specific coupling between the center of mass and rotation.
In \fref{fig:spectrum_energy_vs_omega_repulsive}(d) we show the spectrum for attractively interacting bosons at a rotation frequency $\Omega=1/2$, corresponding  to half of the periodicity expected for non-interacting or repulsively interacting particles. For completeness, in  \fref{fig:spectrum_energy_vs_omega_repulsive}(e) we show an example where a large momenta $\Omega=L/4$ is induced, which corresponds to the a quarter of the maximum angular momenta allowed by the periodicity imposed by the lattice.
In addition, $\Omega$ changes the magnitude of the relative momenta, which is directly related to the characteristics of the bound states, e.g. the decay length in the density-density correlation function. In our lattice system, this coupling between the relative and center of mass momenta (disappearing in the continuous Lieb-Lineger case) has substantial implications on the dynamics of the system. 

A doubling of the periodicity occurs in the lowest energy brand of the spectrum. This behavior, that should be contrasted with the repulsive case, has clear implications on the periodicity of the persistent current \cite{Piero_2019_angluar}.

Figure~\ref{fig:spectrum_energy_vs_omega_repulsive}(e) shows the groundstate energy of the system as a function of the frequency of rotation. Note that the change of parabola corresponds to the degeneracy point between the two lowest energy eigenstates of the system. Contrary to the predictions of the Lieb-Liniger model, the doubling of periodicity in the lattice does depend on the strength of the interactions.

\subsection{Persistent current}
For attractive interactions, the ground state persistent current also displays the characteristic saw-tooth dependence on the effective magnetic field. Compared with the positive $U$ case, the persistent current exhibits a fine structure. 
The current jumps have a clear $N$ dependence. In addition, the slope of the saw-tooth depends on $U$. 
This scenario implies that the flux quantum is reduced by bound states in which more particles can share the same amount of magnetic flux (see \fref{fig:current_vs_omega_fluctuations}(b)), indicating fractionalization of angular momentum per particle \cite{Piero_2019_angluar}. In this case, the composition of the bound states is a variable quantity depending on the interaction.  As a consequence, the response to the magnetic field can be different for different interactions. 

For attractive interactions, the current fluctuations follow the $\Omega$ periodicity of the current. In comparison with the repulsive case,  $\Delta I (\Omega)$ are  much more pronounced. Such distinctive feature is expression of the different impact that the non-conservation of current has for repulsive and attractive interaction:  Going from positive to negative $U$,  multiple occupancy of the lattice sites is more and more probable and therefore  the effect of the  current's non conservation  increases accordingly.

Because of the different nature of  coherence of attracting bosons, the TOF interference fringes display marked differences from those obtained in the repulsive case. For both $\Omega=0$ and $\Omega \neq 0$, we observe a broad peak centered around $\mathbf{k}=0$. Remarkably, the information on the current states is still encoded in the TOF as shown in \fref{fig:TOF_L_11}(d-f). In particular, in \fref{fig:TOF_L_11}(d) we calculate  the central peak $n_k(\mathbf{k} =0)$, which displays clear jumps between different plateaus, directly showing the fractionalization of the angular momentum.



\section{Conclusions}
\label{conclusions}
In this paper  we study the exact ground state properties of the Bose-Hubbard model for two interacting bosons moving in a ring-shaped potential pierced by an effective magnetic field. In this case, the wave function can be expressed as a suitable combination of plane waves  {\it \'a la} coordinate Bethe Ansatz.
Our analysis shows that the interaction can couple the center of mass of the particles and their relative coordinate. This characteristic trait of the lattice system, that is lost in the continuous (integrable) Bose gas or Lieb-Lineger theory, leads to striking consequences in the structure of the ground state, particularly for attractive interactions. In the language of the Bethe Ansatz, while for repulsive interactions the ground state of the system is made of scattering states with real rapidities, for negative $U$ the ground state is a bound state with complex rapidities with a two-string structure. In the latter case, it was demonstrated that the bound states describe the quantum analog of bright solitons \cite{Naldesi2019}. We note that for sufficiently large interactions these bound states are protected by a finite energy gap. This feature is lost in the continuous case. 

We note that the center of mass and relative coordinate coupling has mild consequences on the persistent current for repulsive interactions indicating that  the two  coordinates cannot be resolved in scattering states. In the case of attractive interactions, instead, the $\Omega$ periodicity of the ground state energy and therefore the $\Omega$ dependence of the persistent current is affected by the total number of particles, i.e. doubled in this case (the general $N$ dependence has been studied in  \cite{Piero_2019_angluar}). This effect is a manifestation of the formation of a composite particle made out of $2$-bosons. Because of the non-trivial dynamics of the relative coordinate, a more subtle effect emerges. Indeed, the $\Omega$ periodicity of the ground state energy does depend on the interaction. This scenario indicates that the aforementioned composite object can respond as a particle with a variable mass depending on the interaction. Therefore, the flux quantum for attractive bosons is also a variable quantity that depends on the interaction. 

In a cold atom setting, such effects are visible through the time of flight expansion of the condensate $\langle n(\mathbf{k}) \rangle$. While the current state in repulsive bosons displays the characteristic ring-shape suppression of the density at $\mathbf{k}=0$ in $\langle n(\mathbf{k}) \rangle$, the persistent current of attractive bosons remains peaked at $\mathbf{k}=0$. Despite the seemingly featureless interference, the flux quantum fractionalization emerges as a quantized dispersion of the fringe around the peak at $\mathbf{k}=0$. 

Our results can be used as a benchmark for numerical or other approximated schemes for the many-body problem. In particular, it would be interesting to study how the structure of the ground state is modified by the departure from the integrability.

\bigskip

\begin{acknowledgments}
We thank V. Dunjko, M. Olshanii and H. Perrin for discussions. The Grenoble LANEF framework (ANR-10-LABX-51-01) is acknowledged for its support with mutualized infrastructure. 
We also acknowledge financial support from the ANR project SuperRing (Grant No.  ANR-15-CE30-0012).

\end{acknowledgments}

\bigskip


\appendix


\section{Synthetic gauge field}
Synthetic gauge fields can be generated in different ways \cite{dalibard2011colloquium, dalibard2015introduction}. 
Here, we summarise how synthetic gauge fields can be induced by stirring the condensate.
The rotation of the condensate can be induced by a time-dependent potential $V(x\!-\!\Omega t)$ moving at an angular velocity $\Omega$ on a ring of radius $R$. The time-dependent Hamiltonian is 
\begin{equation}
{\cal H}(\Omega, t) = {\cal H}_0 + V(x\!-\!\Omega R t)	\, , 
\end{equation}
where $ {\cal H}_0 = \sum_{l=0}^N - \frac{\hbar^2}{2m} \nabla_l^2+\hat{U}_{int}$, 
is expressed by the sum of the kinetic energy and a potential term describing the atom-atom interactions $\hat{U}_{int}$. In the following, we assume that $\hat{U}_{int}=U \sum_{l, m}\delta(\mathbf{x}_l-\mathbf{x}_m)$.
By changing to the rotating reference frame with the same frequency $\Omega$ as the external potential, the Hamiltonian is not time dependent: 
\begin{equation}
{\cal H}_{rot} = \mathcal{U}^\dagger(t) \: {\cal H}(\Omega, t) \: \mathcal{U}(t)= {\cal H}_0 + V(x) - \Omega L_z 	\, , 
\end{equation}
where $\mathcal{U}(t) = \text{exp} [ i L_z \Omega t /\hbar ] $, and $L_z$ is the $z$-component angular momentum operator, with $z$ being the coordinate perpendicular to the plane of the ring. 

In second quantization, the Hamiltonian above reads 
\begin{equation}
H=\int d \mathbf{x} \Psi^\dagger (\mathbf x) {\cal H}_{rot} \Psi (\mathbf x) \;, 
\end{equation}
where $\Psi(\mathbf{x})$ are bosonic field operators. 
Using the standard procedure of completing the squares in the kinetic terms, the Hamiltonian reads:
\begin{equation}
H=\int d \mathbf{x} \Psi^\dagger (\mathbf x) \left (\sum_{l=0}^N 
 \frac{(- i\hbar \nabla_l+A)^2}{2m} 
- \frac{A^2}{2m} + V_p \right )\Psi (\mathbf x) \;, 
\end{equation}
where $\vec{A} = R \vec{L}_z $ is the effective vector potential and we define $V_p=+ {U}_{int} + V(x)$.

In a lattice system, the field operators can be expanded in Wannier functions:
$\Psi(\mathbf{x})=\sum_l w_l(\mathbf{x})a_l $, being $a_l$ the site $l$ annihilation bosonic operator. Therefore:
\begin{eqnarray}
H=\sum_{l, m} \int d \mathbf{x} w_l(\mathbf{x}) 
[ \frac{(- i\hbar \sum_l \nabla_l+A)^2}{2m} 
- \frac{A^2}{2m} +\\ \nonumber 
+ V_p ]
 w_m(\mathbf{x}) a^\dagger_l a_m \;.
 \end{eqnarray}

The vector potential can be gauged away by redefining the Wannier functions: $\tilde{w}_l(\mathbf{x}) = w_l(\mathbf{x}) 
e^{iA\mathbf{x_l}} $. This procedure, known as Peierls substitution, leads to our Hamiltonian in Eq.\ref{eq:HBH}. 

\section{Twisted boundary conditions}
By performing a unitary transformation of the Hamiltonian given in \eref{eq:HBH}, in this case a {\it rotation} $U=e^{2\pi i \Omega}$, we change from periodic boundary conditions to twisted boundary conditions \cite{shastry1990twisted, osterloh2000exact}. Such transformation simplifies the systems Hamiltonian to the one obtained for a non-rotating system, i.e. with real tunneling amplitudes. Moreover, we also obtain simplified equations for the energy, center of mass and relative momenta which now read:
\begin{eqnarray}
E^{\text{\tiny TB}}				&=&	-4J\cos\left( \frac{P^{\text{\tiny TB}}}{2} \right)\cos(p^{\text{\tiny TB}}) \\
P^{\text{\tiny TB}}_n 			&=& \frac{2\pi }{L}\left( n -2\Omega \right)\\
(-1)^ne^{ip^{\text{\tiny TB}}L}	&=&	y^{\text{\tiny TB}}\left(P_n, p\right)	\, , 
\end{eqnarray}
with 
\begin{eqnarray}
y^{\text{\tiny TB}}(P, p)			&=&	\frac{U-i4J\cos\left( \frac{P}{2} \right)\sin\left( p \right)}{U+i4J\cos\left( \frac{P}{2} \right)\sin\left( p \right)} 
\end{eqnarray}
Note that in the previous equations, the induced rotation determined by the frequency $\Omega$ only appears explicitly in the center of mass coordinate. Nonetheless, both center of mass and relative coordinates are still coupled as can be seen through the continuity of the energy and implicitly though $y^{\text{\tiny TB}}(P, p)$.

\bibliography{library}

\end{document}